\documentclass[12pt,a4paper]{article}
\usepackage{latexsym,graphicx,psfrag,amsmath}

\newcommand{\bda}{\begin{\displaymath}\begin{array}{rl}}
\newcommand{\eda}{\end{array}\end{displaymath}}
\newcommand{\be}{\begin{equation}}
\newcommand{\ee}{\end{equation}}
\newcommand{\bdm}{\begin{displaymath}}
\newcommand{\edm}{\end{displaymath}}
\newcommand{\bea}{\begin{eqnarray}}
\newcommand{\eea}{\end{eqnarray}}

\newcommand{\fs}{\,.}
\newcommand{\co}{\,,}

\newcommand{\ubar}{\overline{\rule[0.42em]{0.4em}{0em}}\hspace{-0.5em}u}
\newcommand{\dbar}{\,\overline{\rule[0.65em]{0.4em}{0em}}\hspace{-0.6em}d}

\newcommand{\lbar}{\bar{\ell}}
\newcommand{\Kbar}{\,\overline{\rule[0.62em]{0.5em}{0em}}\hspace{-0.8em}K}

\newcommand{\lvac}{\langle 0|\,}
\newcommand{\rvac}{\,|0\rangle}

\newcommand{\al}{&\!\!\!}
\newcommand{\ChPT}{$\chi$PT\hspace{1mm}}
\newcommand{\sA}{s_{\hspace{-0.04cm}A}}
\newcommand{\KPYIII}{KPY\hspace{-0.1cm} III\hspace{0.1cm}}
\def\gn#1{\hspace*{3em} \pageref{#1}}%
\def\bib#1{\vspace*{-0.2cm}\bibitem{#1}}%

\begin{document}
\begin{center}
{\LARGE \bf Physics of the light quarks} \\

\vspace{0.8cm}
H.~Leutwyler, Institute for Theoretical Physics,\\ University of Bern,
Sidlerstr.~5, CH-3012 Bern, Switzerland
\end{center}
\vspace{0.2cm}

\begin{abstract}These lecture notes concern recent developments in our understanding of the low energy properties of QCD. Significant progress has been made on the lattice and the beautiful experimental results on the $K_{e4}$ and $K_{3\pi}$ decays, as well as those on pionic atoms also confirm the results obtained on the basis of Chiral Perturbation Theory. There is an exception: one of the precision experiments on $K_{\mu3}$ decay is in flat contradiction with  the Callan-Treiman relation. If confirmed, this would indicate physics beyond the Standard Model: right-handed quark couplings of the $W$-boson, for instance. Furthermore,  I discuss two examples where the estimates of the effective coupling constants based on saturation by resonances appear to fail.  

In the second part, the progress made in extending the range of validity of the effective theory with dispersive methods is reviewed. In particular, I draw attention to an exact formula, which expresses the mass and width of a resonance in terms of obser\-vable quantities. The formula removes the ambiguities inherent in the analytic continuation from the real axis into the complex plane, which plagued previous determinations of the pole positions associated with broad resonances. In particular, it can now be demonstrated that the lowest resonance of QCD carries the quantum numbers of the vacuum. \end{abstract} 

\vspace{1cm}\begin{center}
Lectures given at the International School of Subnuclear Physics\\  Erice, Italy, 29 August -- 7 September 2007
\end{center}
\newpage
\begin{center}
{\bf \large Contents}

\vspace{0.3cm}
\begin{tabular}{rlr}
1\al Introduction & \gn{sec:intro}\\
2\al Lattice results for $M_\pi$ and $F_\pi$ & \gn{sec:lat}\\
3\al S-wave $\pi\pi$ scattering lengths & \gn{sec:S-wave scattering lengths}\\
4\al Precision experiments at low energy &\gn{sec:Precision experiments}\\
5\al Expansion in powers of $m_s$ & \gn{sec:ms}\\
6\al Violations of the OZI rule ? & \gn{sec:OZI}\\
7\al Problems with scalar meson dominance ? & \gn{sec:scalar meson dominance}\\
8\al Puzzling results in $K_{\mu3}$ decay & \gn{sec:Kmu3}\\
9\al Dispersion theory & \gn{sec:Dispersion theory}\\
10\al Mathematics of the Roy equations & \gn{sec:math}\\
11\al Low energy analysis of $\pi\pi$ scattering  & \gn{sec:Low energy}\\
12\al Behaviour of the S-wave with $I = 0$& \gn{sec:delta00}\\
13\al Pole formula & \gn{sec:Pole formula}\\
14\al The lowest resonance of QCD & \gn{sec:Lowest resonance}\\
\end{tabular}
\end{center}
\setcounter{section}{0}
\section{Introduction}\label{sec:intro}
QCD with massless quarks is the ideal of a theory: it does not contain a
single dimensionless free parameter. At high energies, the degrees of freedom
occurring in the Lagrangian are suitable for a description of the phenomena, because the interaction among these degrees of freedom can be treated as a
perturbation. At low energies, on the other hand, QCD reveals a rich spectrum
of hadrons, the understanding of which is beyond the reach of perturbation
theory. In my opinion, one of the main challenges within the Standard Model is
to understand how an intrinsically simple beauty like QCD can give rise to the
amazing structures observed at low energy.

The progress achieved in understanding the low energy properties of QCD has
been very slow. A large fraction of the papers written in this field does not
concern QCD as such, but models that resemble it in one way or the other:
constituent quarks, NJL-model, linear $\sigma$ model, hidden local symmetry,
AdS/CFT and many others. Some of these may be viewed as simplified versions of
QCD that do catch some of the salient features of the theory at the
semi-quantitative level, but none provides a basis for a coherent
approximation scheme that would allow us, in principle, to solve QCD.
 
These lectures concern the model independent approach to the problem based on
effective field theory, lattice methods and dispersion theory. The effective theory relevant for low energy QCD is referred to as Chiral Perturbation Theory, \ChPT. For recent reviews of this framework in the mesonic sector, I refer to \cite{Bijnens}--\cite{Colangelo Erice}. An update of the experimental information concerning the effective coupling constants was given by Bijnens at the lattice conference in 2007 \cite{Bijnens lattice07}. The rapidly growing information obtained in the framework of the lattice approach is reviewed in the report of Necco at the same meeting \cite{Necco lattice07}.  An up-to-date account of the progress made in applying \ChPT to the baryons is given in \cite{Veronique Bernard}.

At low energies, the main characteristic of QCD is that the energy gap is remarkably
small, $M_\pi\simeq $ 140 MeV. More than 10 years before the discovery of QCD,
Nambu \cite{Nambu} found out why that is so: the gap is small because the
strong interactions have an approximate chiral symmetry. Indeed, QCD does have
this property: for yet unknown reasons, two of the quarks happen to
be very light. The symmetry is not perfect, but nearly so: $m_u$ and $m_d$ are
tiny. The mass gap is small because the symmetry is ``hidden'' or
``spontaneously broken'': for dynamical reasons, the ground state of the
theory is not invariant under chiral rotations, not even approximately. The
spontaneous breakdown of an exact Lie group symmetry gives rise to strictly
massless particles, ``Goldstone bosons''. In QCD, the pions play this role:
they would be strictly massless if $m_u$ and $m_d$ were zero, because the
symmetry would then be exact. The only term in the Lagrangian of QCD that is not
invariant under the group SU(2)$\times$SU(2) of chiral rotations
is the mass term of the two lightest quarks, $m_u\,\ubar u+m_d \,\dbar
d$. This term equips the pions with a mass. Although the
theoretical understanding of the ground state is still poor, we do have very
strong indirect evidence that Nambu's conjecture is right -- we know why the
energy gap of QCD is small.

\section{Lattice results for $M_\pi$ and $F_\pi$}\label{sec:lat}
As pointed out by Gell-Mann, Oakes and Renner \cite{GMOR}, the square of the
pion mass is proportional to the strength of the symmetry breaking,
\bdm M_\pi^2\propto (m_u+m_d)\,.\edm This property can now be checked on the lattice,
where -- in principle -- the quark masses can be varied at will. In
view of the fact that in these calculations, the quarks are treated
dynamically, the quality of the data is impressive. The masses are
sufficiently light for \ChPT to allow a meaningful extrapolation to the
quark mass values of physical interest. The results indicate that the ratio
$M_\pi^2/(m_u+m_d)$ is nearly constant out to values of $m_u, m_d$ that are
about an order of magnitude larger than in nature. 

The Gell-Mann-Oakes-Renner relation corresponds to the leading term in the
expansion in powers of the quark masses. At next-to-leading order, the
expansion in powers of $m_u,m_d$ (mass of the strange quark kept fixed at the physical value) contains a logarithm \cite{Langacker and Pagels, Gasser 1981}: 
\be\label{eq:Mpi one loop} M_\pi^2=
M^2\left\{1 +\!\frac{M^2}{32\pi^2 F_\pi^2}\, \ln
  \frac{M^2}{\Lambda_3^2}\!+\!O(M^4)\right\}\co\ee
where $M^2\equiv B(m_u+m_d)$ stands for the term linear in the quark masses. 
Chiral symmetry fixes the coefficient of the logarithm in
terms of the pion decay constant $F_\pi$, but does not determine the scale
$\Lambda_3$. A crude estimate for this scale was obtained more than 20 years
ago \cite{GL SU2}, on the basis of the SU(3) mass formulae for the pseudoscalar
octet. The result is indicated at the bottom of the left panel in figure 1. 
\begin{figure}[thb] 
\parbox{15cm}{\psfrag{l3bar}{\hspace{-0.8cm}\raisebox{-0.5cm}
{$\displaystyle\bar{\ell}_3=\ln\frac{\Lambda_3^{\;2}}{M_\pi^2}$}}
\psfrag{l4bar}{\hspace{-1cm}\raisebox{-0.5cm}{$\displaystyle\bar{\ell}_4=\ln
    \frac{\Lambda_4^{\;2}}{M_\pi^2}$}} 
\includegraphics[height=.33\textheight,angle=-90]{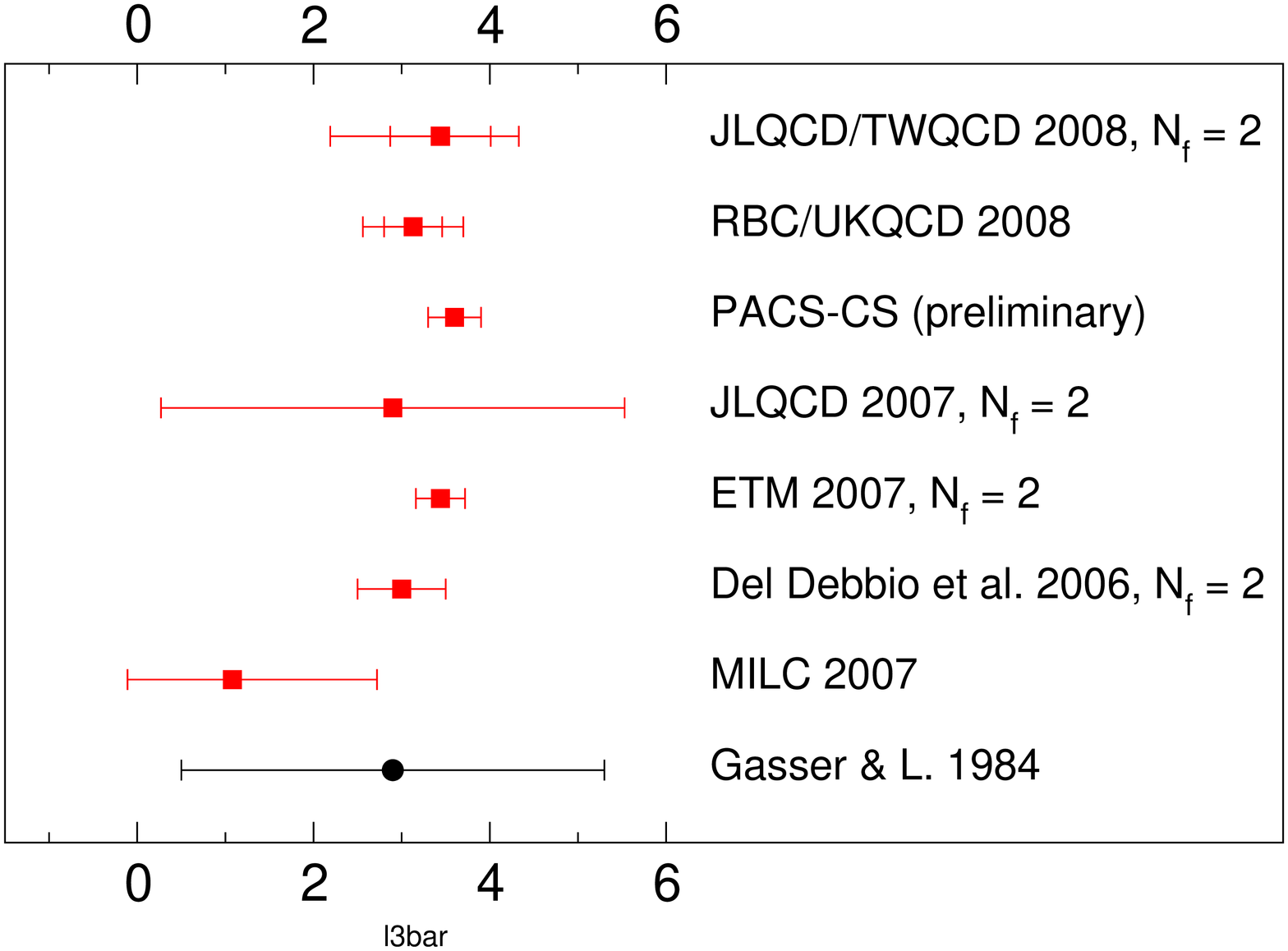}
\includegraphics[height=.33\textheight,angle=-90]{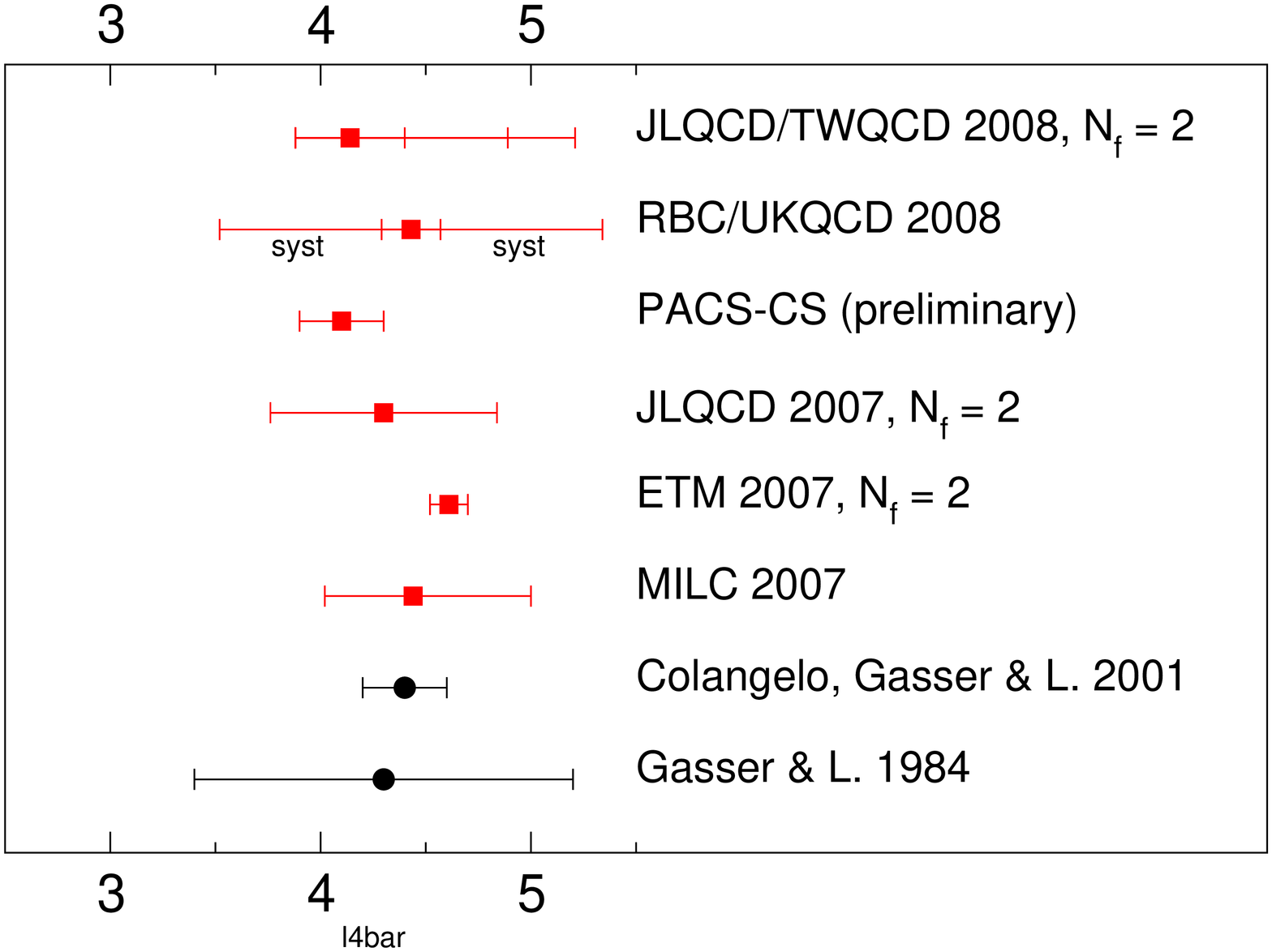}\\
\rule{0cm}{0.8cm}}

\caption{\label{fig:l3l4}Determinations of the effective coupling
  constants $\ell_3$ and $\ell_4$}  
\end{figure}
The other entries represent recent lattice results for this quantity \cite{MILC}--\cite{JLQCD/TWQCD}. The one of the RBC /UKQCD collaboration, $\lbar_3=3.13 \pm 0.33_{\,\mbox{\tiny stat}}\pm 0.24_{\,\mbox{\tiny syst}}$ \cite{RBC/UKQCD}, for instance, which concerns 2+1 flavours and includes an estimate of the systematic errors, is considerably more accurate than our old estimate based on SU(3), $\lbar_3=2.9\pm 2.4$ \cite{GL SU2}.

The right panel shows the results for the scale $\Lambda_4$, which determines the quark mass dependence of the
pion decay constant at NLO of the chiral expansion. The analog of formula (\ref{eq:Mpi one loop}) reads 
\be\label{Fpi one loop} F_\pi=
F\left\{1 -\!\frac{M^2}{16\pi^2 F^2}\, \ln
  \frac{M^2}{\Lambda_4^2}\!+\!O(M^4)\right\}\co \ee 
where $F$ is the value of the pion decay constant in the limit $m_u,m_d\rightarrow 0$. A couple of years ago, we obtained a rather accurate result for $\Lambda_4$, from a dispersive analysis of the scalar pion form factor \cite{CGL}.  The plot shows that the lattice determinations of $\bar{\ell}_4$ have reached comparable accuracy and are consistent with the dispersive result. 
 For a detailed discussion of the properties of the scalar pion form factor, I refer to \cite{ACCGL}. This quantity is now also accessible to an evaluation on the lattice \cite{Kaneko}.

\section{S-wave $\pi\pi$  scattering lengths}
\label{sec:S-wave scattering lengths}
The hidden symmetry not only controls the size of the energy gap, but also
determines the interaction of the Goldstone bosons at low energies, among
themselves, as well as with other hadrons. In particular, as pointed out by
Weinberg \cite{Weinberg 1966}, the leading term in the chiral expansion of the
S-wave $\pi\pi$ scattering lengths (tree level of the effective theory) is
determined by the pion decay constant. The corresponding numerical values of
$a_0^0$ and $a_0^2$ are indicated by the leftmost dot in figure 2, while the
other two show the result obtained at NLO and NNLO of the chiral expansion,
respectively. The exotic scattering length $a_0^2$ is barely affected by
the higher order corrections, but the shift seen in $a_0^0$ is quite
substantial. The physics behind this enhancement of the perturbations
generated by $m_u$ and $m_d$ is well understood: it is a consequence of the
final state interaction, which is attractive in the $S^0$-wave, rapidly grows
with the energy and hence produces chiral logarithms with large coefficients. 

\begin{figure}[thb] \parbox{15cm}{
    \includegraphics[height=.34\textheight,angle=-90]{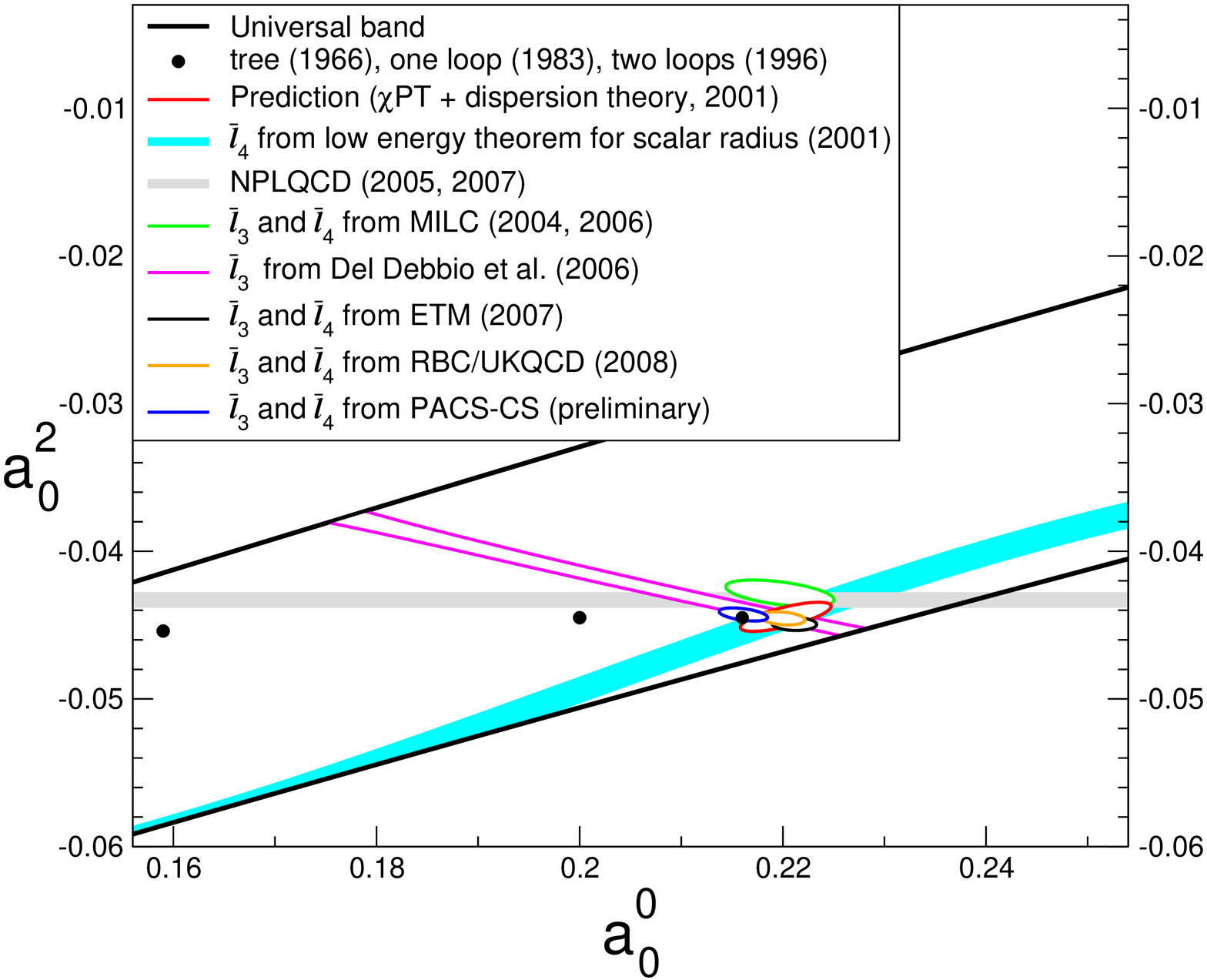}
    \includegraphics[height=.34\textheight,angle=-90]{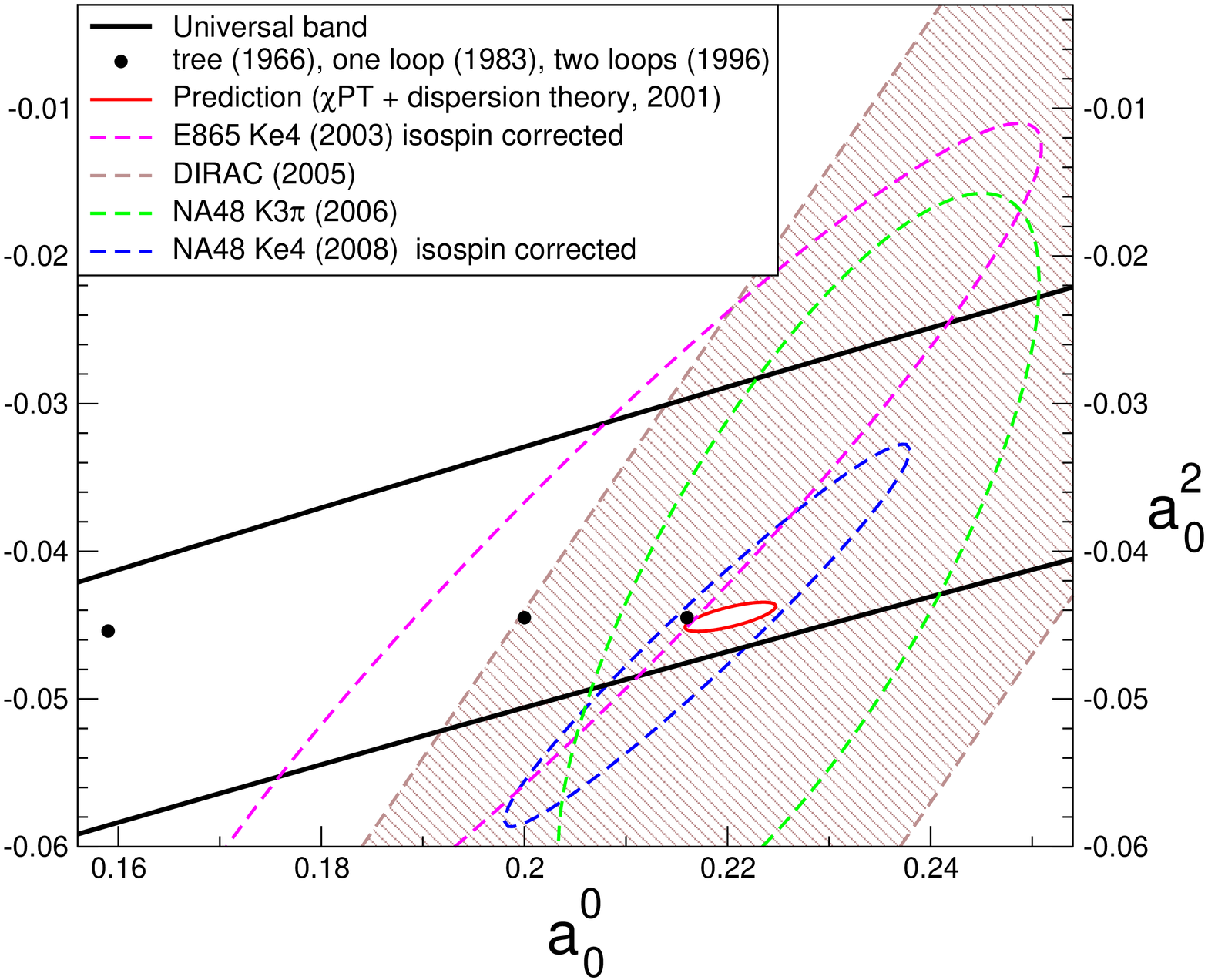}\\
    \rule{0cm}{0cm}}
\caption{Comparing the theoretical predictions for the $\pi\pi$ S-wave
  scattering lengths with lattice results (left) and with experiment (right)} 
\end{figure}
Near the center of the Mandelstam triangle, the
contributions from higher orders of the chiral expansion are small \cite{CGL}. Using
dispersion theory to reach the physical region from there, we arrived at the remarkably
sharp predictions for the two scattering lengths indicated on the left panel
of figure 2. Our analysis also shows that the corrections to Weinberg's low
energy theorem for $a_0^0, a_0^2$ are dominated by the effective coupling
constants $\bar{\ell}_3,\bar{\ell}_4$ discussed above -- if these are known,
the scattering lengths can be calculated within small uncertainties. Except
for the horizontal band, which represents a direct determination of $a_0^2$
based on the volume dependence of the levels \cite{NPLQCD}, all of the lattice results for the scattering lengths shown on the left panel of figure 2 are obtained in this
way from the corresponding results for $\ell_3$ and $\ell_4$. The figure
demonstrates that the lattice results confirm the predictions for
$a_0^0,a_0^2$. 

\section{Precision experiments at low energy}\label{sec:Precision experiments}
The right panel of figure 2 compares the predictions for the scattering lengths
with recent experimental results. While the $K_{e4}$ data of E865 \cite{E865},
the DIRAC experiment\cite{DIRAC} and the NA48 data on the cusp in
$K\rightarrow 3\pi$ \cite{NA48 cusp} all confirm the theoretical expectations,
the most precise source of information, the beautiful $K_{e4}$ data of NA48
\cite{NA48 paper on Ke4}, gave rise to a puzzle. The Watson theorem implies
that -- if the electromagnetic interaction and the difference between $m_u$
and $m_d$ are neglected -- the relative phase of the form factors describing
the decay $K\rightarrow e\nu\pi\pi$ coincides with the difference
$\delta_0^0-\delta_1^1$ of scattering phase shifts. At the precision achieved,
the data on the form factor phase did not agree with the theoretical prediction
for the phase shifts. 
\begin{figure}[thb]\centering
\includegraphics[height=.4\textheight,angle=-90]{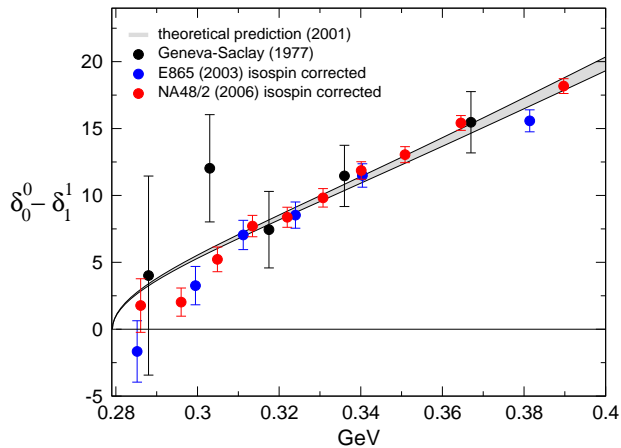}
\caption{Comparison of $K_{e4}$ data with the prediction for
  $\delta^0_0-\delta^1_1$}  
\end{figure} 

The origin of the discrepancy was identified by Colangelo, Gasser and Rusetsky
\cite{Colangelo Gasser Rusetsky}. The problem has to do with the fact that
a $K^+$ may first decay into $e^+\nu\hspace{0.5mm}\pi^0\pi^0$ -- the pair of neutral pions then undergoes scattering and winds up as a charged pair. The mass difference
between the charged and neutral pions affects this process in a pronounced
manner: it pushes the form factor phase up by about half a degree -- an
isospin breaking effect, due almost exclusively to the electromagnetic
interaction.

Figure 3 shows that the discrepancy disappears if the NA48 data
on the relative phase of the form factors are corrected for isospin breaking.
Accordingly, the range of scattering lengths allowed by these data, shown on
the right panel of figure 2, is in perfect agreement with the prediction. As indicated on the left panel of that figure, the low energy theorem for the scalar radius of the pion correlates the two S-wave scattering lengths to a narrow strip. If this correlation is used, the analysis of the $K_{e4}$ data leads to $a_0^0=0.220 \pm 0.005_{\,\mbox{\tiny stat}}\pm 0.002_{\,\mbox{\tiny syst}}$ \cite{Bloch-Devaux Anacapri 2008}. The result has the same precision as the theoretical prediction and hits it on the nail.  I conclude that the puzzle is gone: $K_{e4}$ confirms the
theory to remarkable precision.

 \section{\boldmath Expansion in powers of $m_s$}\label{sec:ms}
The examples discussed above all concern the effective theory based on SU(2)$\times$SU(2), where the quantities of interest are expanded in powers of $m_u,m_d$, while $m_s$ is kept fixed at the physical value. The corresponding effective coupling constants $F,B,\ell_1,\ldots$ are independent of $m_u$ and $m_d$, but do depend on $m_s$. Their expansion in powers of $m_s$ can be worked out in the framework of the effective theory based on SU(3)$\times$SU(3).  For $F,B$, the expansion starts with\footnote{The contributions of $O(m_s^2)$ are also known explicitly, not only for $F,B$, but also for the coupling constants $\ell_1,\ldots,\ell_7$, which specify the effective Lagrangian at NLO  \cite{GHIS,Kaiser Mpi}.}
 \bea\label{eq:FB} F\al=\al F_0\left\{ 1+\frac{8\bar{M}^2_K}{F_0^2}L_4^r-\bar{\mu}_K+O(m_s^2)\right\}\,,\\
B\al=\al B_0\left\{ 1+\frac{16\bar{M}^2_K}{F_0^2}(2L_6^r-L_4^r)-\bar{\mu}_\eta+O(m_s^2)\right\}\,.\nonumber\eea
The constants $F_0,B_0$ represent the values of $F,B$ in the limit $m_s\rightarrow 0$. At NLO, only the coupling constants $L_4,L_6$ of the chiral SU(3)$\times$SU(3) Lagrangian enter, weighted with the square of the kaon mass in the limit $m_u,m_d=0$, which I denote by $\bar{M}_K$. In this limit, the octet of Goldstone bosons contains only three different mass values: $\bar{M}_\pi=0$,  $\bar{M}_K$ and $\bar{M}_\eta$. To the accuracy relevant in the above formulae, we have $\bar{M}_K^2=B_0\,m_s$, $\bar{M}_\eta^2=\frac{4}{3}B_0\,m_s$. Up to corrections of higher order, $\bar{M}_K$ and $\bar{M}_\eta$ may be expressed in terms of the physical masses as
\be \bar{M}_K^2=M_K^2-\mbox{$\frac{1}{2}$}M_\pi^2\,,\hspace{1cm}
\bar{M}_\eta^2=\mbox{$\frac{4}{3}$}M_K^2-\mbox{$\frac{2}{3}$}M_\pi^2\fs\ee
The chiral logarithms occurring in the above formulae may be expressed in terms of the function
\be \bar{\mu}_P=\frac{\bar{M}_P^2}{32 \pi^2 F_0^2}\ln\frac{\bar{M}_P^2}{\mu^2}\,,\hspace{1cm}P=K,\eta\,.\ee
They involve the running scale $\mu$ at which the chiral perturbation series is renormalized, but the scale dependence of the renormalized coupling constants $L_4^r,L_6^r$ ensures that the expressions in the curly brackets of equation (\ref{eq:FB}) are scale independent.

The quark condensate, 
\be \Sigma =| \lvac \ubar u\rvac|\rule[-2mm]{0mm}{5mm}_{\;m_u,m_d\rightarrow 0}\co\ee
is determined by the same two constants:  $\Sigma = F^2 B$. Accordingly, the expansion of the condensate in powers of $m_s$ starts with
\bea\label{eq:Sigma}\Sigma\al=\al \Sigma_0\left\{1+\frac{32\bar{M}^2_K}{F_0^2}L_6^r-2\mu_{\Kbar}-\mu_{\bar{\eta}}+O(m_s^2)\right\}\,.\eea
The coupling constants $L_4, L_6$ as well as the loop graphs responsible for the chiral logarithms represent effects that violate the Okubo-Zweig-Iizuka rule. In the large $N_c$ limit, the quantities $F,B,\Sigma$ become independent of $m_s$, so that the ratios $F/F_0, B/B_0, \Sigma/\Sigma_0$ tend to 1. If the OZI rule is a good guide in the present context, then these ratios should not differ much from 1. For a discussion of the implications of large OZI violations in these ratios, see \cite{Descotes lattice07}. The paramagnetic inequalities of Stern et al.\ \cite{paramagnetic} indicate that the sign of the deviations $F/F_0-1$ and $\Sigma/\Sigma_0-1$ is positive. 

\section{Violations of the OZI rule ?}\label{sec:OZI}
\begin{figure}[thb]\hspace{1.5cm}
\includegraphics[width=6.2cm,angle=-90]{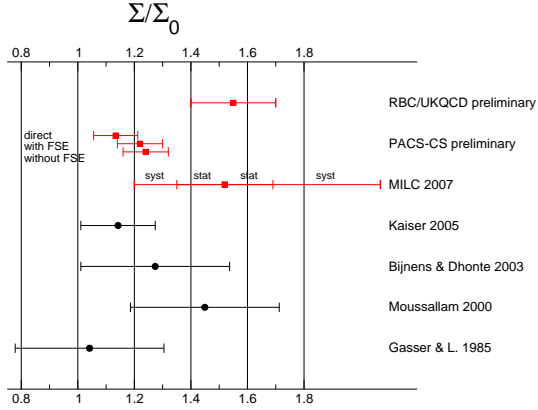} 
\caption{\label{fig:Sigma}Violation of the OZI rule in the quark condensate}\end{figure}
Figure \ref{fig:Sigma} compares recent lattice results for the dependence of the condensate on $m_s$ \cite{MILC,PACS,RBC/UKQCD Lattice 07} with phenomenological estimates found in the literature \cite{GL SU3,Moussallam 2000,Bijnens & Dhonte 2003,Kaiser Trento & Montpellier}. The latter are calculated from the values quoted for the running coupling constant $L_6$, using the relation (\ref{eq:Sigma}) with $F_0=F_\pi$. The errors shown exclusively account for the quoted uncertainties in the coupling constants, while those arising from the corrections of $O(m_s^2)$ are neglected (accounting for the difference between $F_0$ and $F_\pi$ in the curly bracket of (\ref{eq:Sigma}), for instance, generates corrections of this type). The plot shows that the uncertainties in the phenomenological estimates are large. Unfortunately, the lattice results are not yet conclusive, either. Note that some of these are preliminary and do not include an estimate of the systematic errors. 

\begin{figure}[here]
\hspace{-1cm}\includegraphics[width=6.2cm,angle=-90]{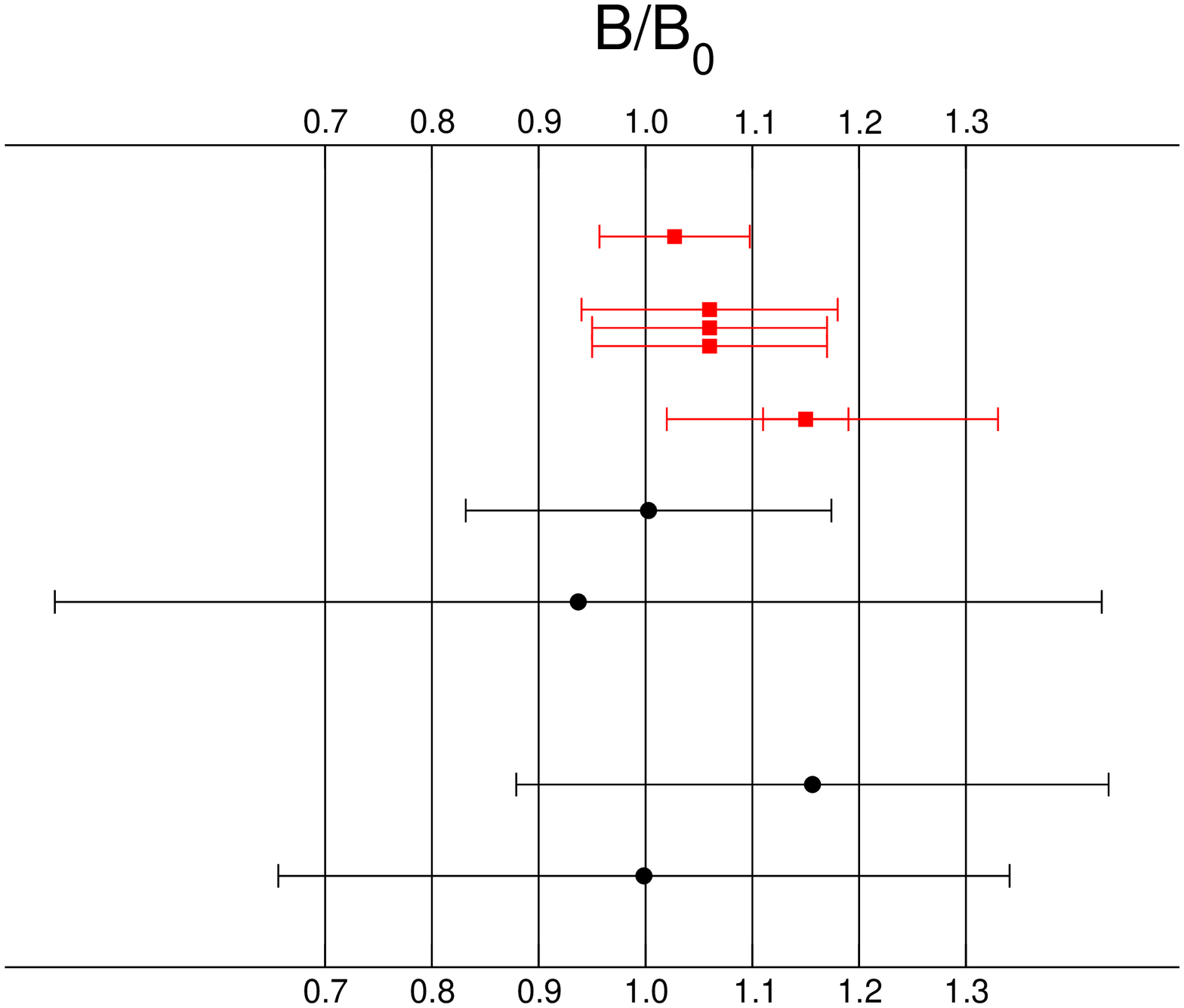}\hspace{-1.2cm}\includegraphics[width=6.2cm,angle=-90]{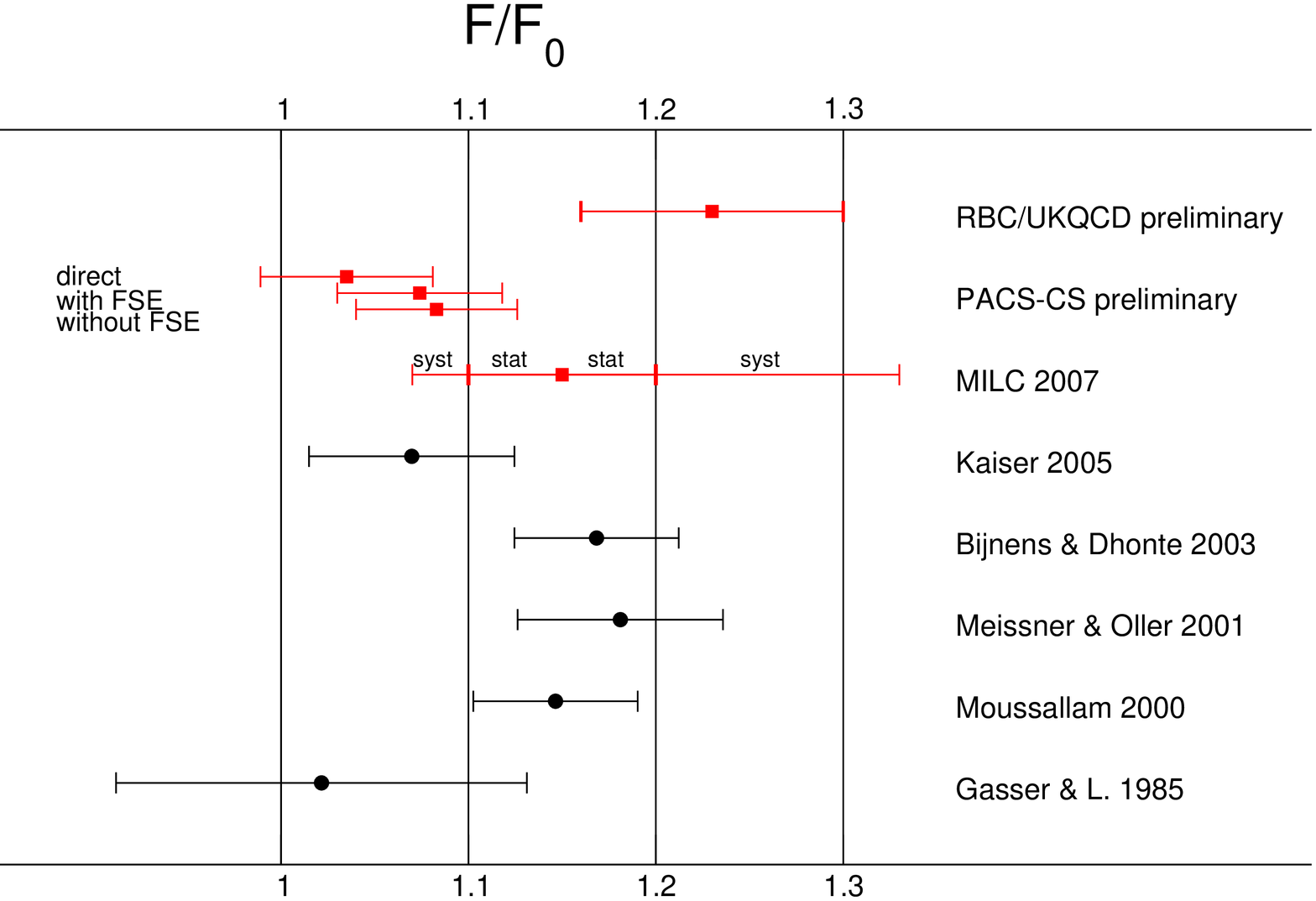}
\caption{\label{fig:BF}OZI violations in the leading effective coupling constants}\end{figure}

In contrast to the lattice results for the condensate, those for the constant $B$ are consistent with one another. As can be seen in figure \ref{fig:BF}, 
the values obtained for $B/B_0$ do not indicate a large violation of the OZI rule. This implies that the discrepancy seen in the lattice results for $\Sigma=F^2B$ originates in the factor $F^2$. Indeed, the values quoted for $F/F_0$ in \cite{RBC/UKQCD Lattice 07} are puzzling, for the following reason. The quantity $F_\pi$ represents the pion wave function at the origin. The value of $F_K$ is somewhat larger, because one of the two valence quarks is heavier than in the case of the pion. Hence it moves mores slowly, so that the wave function is more narrow and thus higher at the origin:\footnote{Note that, a few years ago, improved measurements performed at several laboratories demonstrated that the $K_{\ell3}$ decay rate had been underestimated by more than 3 $\sigma$. As a consequence, the values obtained for the CKM matrix element $V_{us}$ were too small, so that the ratio $F_K/F_\pi$ of decay constants came out too large \cite{Antonelli}. The value quoted  in Eq.\ (\ref{eq:FK/Fpi}) is taken from Bernard and Passemar \cite{Bernard & Passemar}. The mini-review of Rosner and Stone \cite{Rosner & Stone}, prepared for the 2008 edition of the PDG tables confirms this result within errors.}  
\be\label{eq:FK/Fpi} F_K/F_\pi= 1.192(7) \fs\ee If the value of $F/F_0$ was larger than this, we would have to conclude that the wave function is more sensitive to the mass of the sea quarks than to the mass of the valence quarks. I do not see a way to rule this logical possibility out, but it is counter intuitive and hence puzzling.  

For the time being, the only conclusion to draw is that the lattice results confirm the paramagnetic inequalities and indicate that the constant $B$ -- the leading term in the expansion of $M_\pi^2$ in powers of $m_u$ and $m_d$ -- does obey the OZI rule. Some of the data indicate that this rule approximately holds also in the case of $F/F_0$, but others suggest rather juicy violations in that case. The slow, but steady progress being made on the lattice gives rise to the hope that the dust will settle soon.

\section{Problems with scalar meson dominance ?}\label{sec:scalar meson dominance}
The analysis of the lattice data on the quark mass dependence of $M_\pi$, $M_K$, $F_\pi$, $F_K$ in terms of the \ChPT representation to two loops will also make it possible to determine those coupling constants of the effective NNLO Lagrangian that contribute to   these quantities. The theoretical estimates for those couplings \cite{Bijnens}--\cite{Bijnens lattice07}, \cite{resonance estimates} rely on the assumption that the relevant sum rules are saturated by the lowest resonances. I know of  two cases, where calculations within this framework run into a problem:
\begin{itemize}\item The data on nuclear $\beta$ decay lead to a remarkably accurate value  for the element $V_{ud}$ of the CKM matrix: $V_{ud}=0.97418 (26)$ \cite{Towner & Hardy}. The unitarity of this matrix then implies $V_{us}= 0.2258(11)$. Since the  data on the rate of the semileptonic $K$ decays require $f_+^{K^0\pi^-}\hspace{-0.1cm}(0)V_{us}=0.21661(47)$  \cite{Flavianet KAON},  the value of the form factor at the origin is determined very sharply:  $f_+^{K^0\pi^-}\hspace{-0.1cm}(0)=0.9594 (49)$. The recent lattice result, $f_+^{K^0\pi^-}\hspace{-0.1cm}(0)=0.964(5)$ \cite{RBC/UKQCD f+}, as well as those for the ratio $F_K/F_\pi$  \cite{f+ lattice}, which offer an independent determination of $V_{us}$, are consistent with this value.
The chiral representation of the relevant form factors was worked out to NNLO by Bijnens and Talavera \cite{Bijnens & Talavera}.  The evaluation of their representation for $f_+^{K^0\pi^-}\hspace{-0.1cm}(0)$ with resonance estimates for the effective coupling constants \cite{Bijnens & Talavera,Jamin Oller & Pich,Cirigliano Kmu3} leads to values that are significantly higher than those above: the recent update of the calculation described in \cite{Kastner & Neufeld} leads to $f_0^{K^0\pi^-}\hspace{-0.1cm}(0)=0.986\pm 0.007_{1/N_c}\pm 0.002_{M_S,M_P}$.  
\item The branching ratios of the transitions $K^\pm\rightarrow\pi^\pm\pi^0$, $K^0\rightarrow\pi^+\pi^-$, $K^0\rightarrow\pi^0\pi^0$ are affected by the final state interaction. Conversely, the observed values of these ratios can be used to determine the phase difference between the two S-waves at $s=M_K^2$. In the past, work on this problem invariably led to a value for the phase difference that is too large, presumably because isospin breaking, which plays a crucial role here, was not properly accounted for. Only rather recently, Cirigliano, Ecker, Neufeld and Pich have performed a complete analysis of these transitions, based on \ChPT to NLO  \cite{Cirigliano Kpipi}. Their calculation accounts for isospin breaking, both from $m_u\neq m_d$ and from the electromagnetic interaction \cite{Wolfe & Maltman}. Unfortunately, however, the discrepancy persists: their result is $\delta_0^0(M_K^2)-\delta_0^2(M_K^2)=57.5^\circ\pm 3.4^\circ$ \cite{Flavianet Kpipi}, while the Roy analysis of $\pi\pi$ scattering implies\footnote{A comparison with more recent work on the $\pi\pi$ phase shifts is made in section \ref{sec:Low energy analysis}.} $\delta_0^0(M_K^2)-\delta_0^2(M_K^2)=47.7^\circ\pm 1.5^\circ$ \cite{CGL}.\end{itemize} 

The above discussion of $f_+^{K^0\pi^-}\hspace{-0.1cm}(0)$ assumes that, at the accuracy of interest, $K_{\ell3}$ decay is properly described by the Standard Model, where the CKM matrix is unitary. Also, it relies on the value of $V_{ud}$ extracted from nuclear $\beta$ decay. These ingredients should not be taken for granted \cite{thank  Gerhard},\footnote{The results obtained from preliminary data on $\tau$ decays into final states with strange\-ness, for instance, do not agree with the above value of $V_{us}$ \cite{Vus tau}. Also, a recent experiment on neutron decay \cite{Serebrov} came up with a neutron lifetime that strongly disagrees with the world average. If confirmed, this calls for an increase in the value of $g_A/g_V$ or in the value of $V_{ud}$. For the time being, the uncertainties in $g_A/g_V$ are too large for neutron decay to compete with the superallowed nuclear transitions. In particular, the value of the neutron lifetime reported in \cite{Serebrov} is consistent with  the value of $V_{ud}$ obtained in \cite{Towner & Hardy}.}  but I consider it more likely that the discrepancy originates in the chiral calculation, also in the case of the phase difference  $\delta_0^0(M_K^2)-\delta_0^2(M_K^2)$. I do not doubt the chiral representations of the form factors \cite{Bijnens & Talavera} and of the phase difference \cite{Cirigliano Kpipi}, but the estimates used for the effective coupling constants play an equally important role. In my opinion, this is the weakest point in the above two applications of \ChPT. 

In the case of SU(2),  the size of the coupling constants occurring in the NLO effective Lagrangian can be understood on the basis of vector meson dominance alone\footnote{The more sophisticated treatment of the vector meson dominance hypothesis described in \cite{Dominguez} confirms the crude estimate for $\bar{\ell}_4$ given in equations (C.12), (C.13) of \cite{GL SU2}.}  \cite{GL SU2}. The above two examples, however, concern SU(3), where $m_s$ is also treated as a perturbation. In this case, the effects generated by the quark mass term in the Lagrangian of QCD are much more important. Since this term is a scalar operator, resonance estimates for those coupling constants that describe the dependence of the effective SU(3) Lagrangian on the quark masses rely on the scalar meson dominance hypothesis. In my opinion, it is questionable whether the complex low energy structure of the scalar states (strong continuum related to the rapidly rising  $\pi\pi$ interaction, broad bump associated with the $\sigma$, narrow peak from the $f_0(980)$, glueballs, etc.) can adequately be accounted for with this hypothesis. 

As shown in \cite{EGPdeR}, the phenomenological results  \cite{GL SU3} for the coupling constants occurring in the SU(3) Lagrangian at NLO can be understood on the basis of the assumption that the contributions from the lowest resonances with spin $\leq 1$ dominate. At first sight, this might appear to confirm the validity of scalar meson dominance in the present context, but that is not the case: in \cite{EGPdeR}, the quark mass dependence of the effective Lagrangian is taken from phenomenology (the model used for the scalar resonances contains two free parameters and these are tuned so as to reproduce the observed values of $L_5$ and $L_8$). Phenomenological information about the quark mass dependence of the NLO couplings relevant for $K\rightarrow\pi\pi$ is lacking. In this case, factorization is used to estimate the effective coupling constants \cite{thank Cirigliano}. The contributions generated by a single resonance do factorize, but in the more complex situation encountered in the scalar channel, factorization may fail.

In the first example, the available experimental and theoretical information should suffice to determine all of the coupling constants relevant for the form factors to NNLO. Those accounting for the slope and the curvature were worked out already \cite{Jamin Oller & Pich}. It would be very instructive to determine the remaining ones, which describe the dependence of the form factors on the quark masses. Inserting the values obtained in this way in the chiral representation of the form factors should lead to a coherent picture, but the results obtained for some of the coupling constants will necessarily differ from the resonance estimates. It would be most useful to understand the origin of the difference, because similar departures from resonance saturation must be expected in other matrix elements.  Whether this will lead to a resolution of the second discrepancy remains to be seen -- at any rate, the puzzle with the phase determination via $K\rightarrow \pi\pi$ cries for a resolution.

\section{\boldmath Puzzling results in $K_{\mu3}$ decay}\label{sec:Kmu3}
The low energy theorem of Callan and Treiman \cite{Callan Treiman} predicts
the size of the scalar form factor of the decay $K\rightarrow \pi\mu\nu$ at
one particular value of the momentum transfer, namely $t=M_K^2-M_\pi^2$:
\be\label{CT} f_0(M_K^2-M_\pi^2)=\frac{F_K}{F_\pi}+O(m_u,m_d) \fs\ee Within
QCD, the relation becomes exact if the quark masses $m_u$ and $m_d$ are set
equal to zero. The corrections of first nonleading order, which have been
evaluated long ago \cite{GL form factors}, are tiny: they lower the right hand 
side by $3.5\times 10^{-3}$. In the meantime, the chiral perturbation series of
$f_0(t)$ has been worked out to NNLO \cite{Bijnens & Talavera,Post & Schilcher}. As pointed out by Jamin,
Oller and Pich \cite{Jamin Oller & Pich}, the curvature of the form factor
can be calculated with dispersion theory. Their dispersive representation agrees very well with the more recent one of Bernard, Oertel, Passemar and Stern \cite{Stern Kmu3}: theory reliably determines the curvature of the form factor. Accordingly, the theoretical prediction for the value at the Callan-Treiman point, $t=M_K^2- M_\pi^2$, can be converted into a prediction for the slope. The result obtained by Jamin et al.\ in 2004 was $\lambda_0 = 0.016(1)$. The update of their calculation with the improved information available in 2006 led to $\lambda_0= 0.0147(4)$. Within the remarkably small errors, this agrees with the outcome of the recent analysis described in \cite{Bernard & Passemar}, for which the Standard Model prediction is $\lambda_0=0.0150(8)$ \cite{Passemar private comm}.  

Recently, the NA48 collaboration published their results for the
$K^0_{\mu3}$ form factors \cite{NA48 Kmu3}. Their result for the scalar slope,
$\lambda_0=0.0117(7)(1)$, is in flat contradiction with the theoretical prediction just discussed. 

\begin{figure}\centering
\includegraphics[width=7cm,angle=-90]{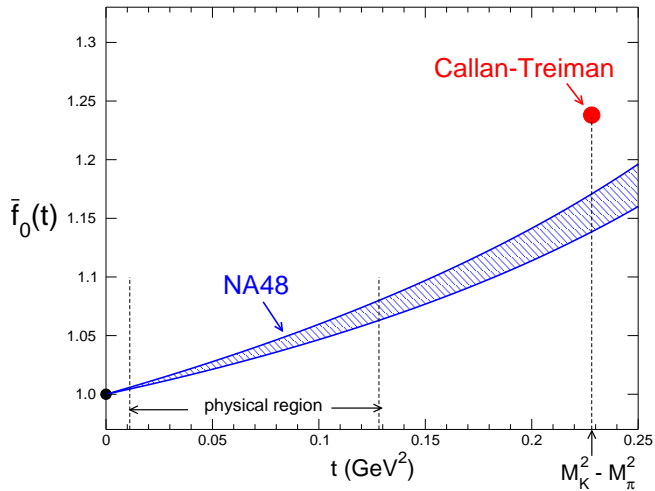}
\caption{NA48 results for the scalar $K_{\mu3}$ form factor}\end{figure}
 
\begin{figure}[thb]\centering
\includegraphics[width=8cm,angle=-90]{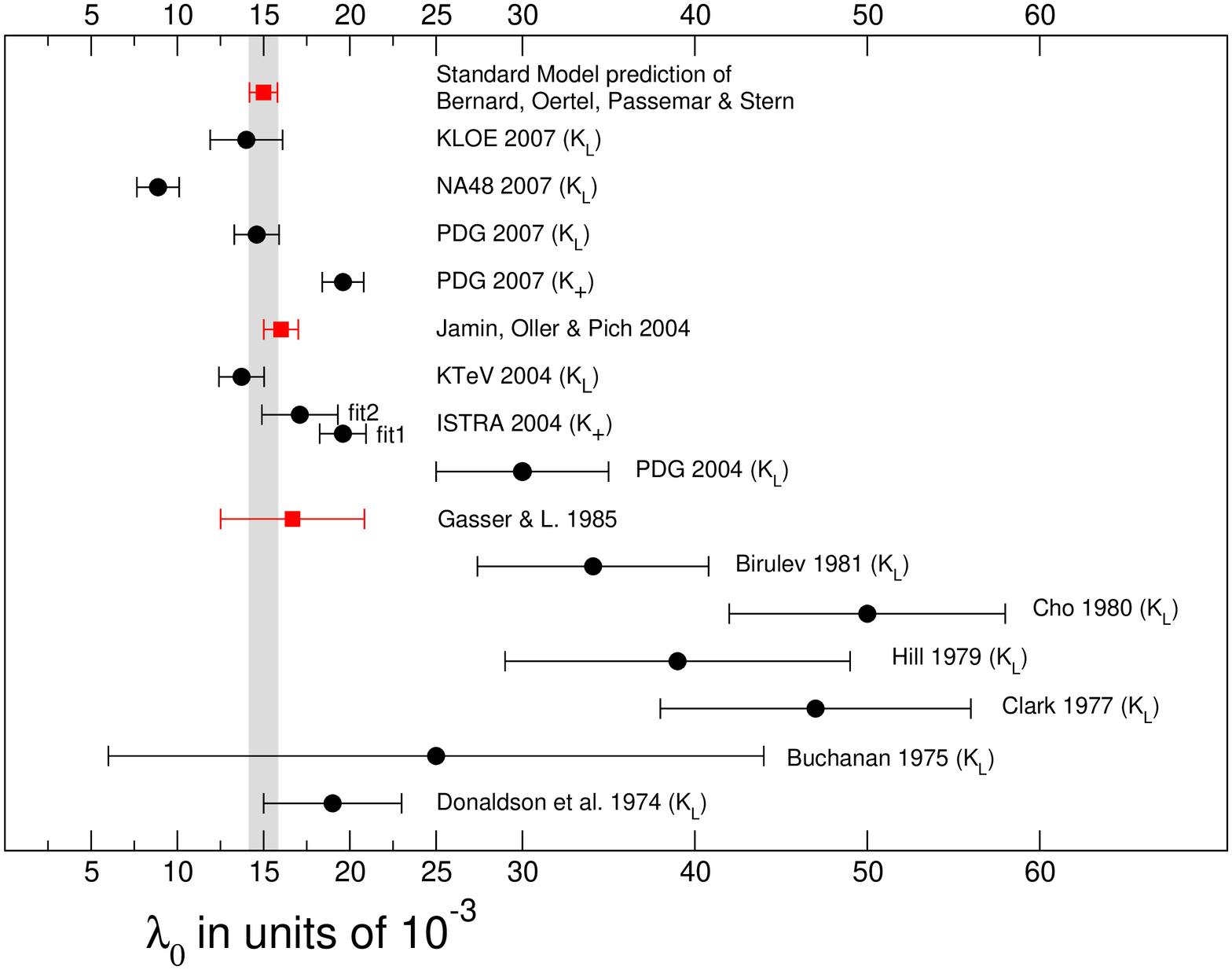}
\caption{\label{fig:lambda0}Old and new results for the slope of the scalar form factor}\end{figure}

The NA48 experiment is not the first to measure the slope of the scalar form
factor relevant for $K_{\ell3}$ decay. Figure \ref{fig:lambda0} compares the outcome of this experiment with results obtained by ISTRA \cite{ISTRA}, KTeV \cite{KTeV}, KLOE \cite{KLOE Kmu3 2007} and with earlier findings, taken from the PDG listings of 2004 \cite{PDG 2004}. The earliest result shown indicates the outcome of the high statistics experiment of Donaldson et al.\ \cite{Donaldson}, which came up with a slope of $\lambda_0=0.019(4)$ and thereby confirmed the validity of the Callan-Treiman relation. The plot shows that quite a few of the experiments performed since then obtained quite different results. 

In 1985, when we worked out the corrections to the Callan-Treiman relation to one loop of \ChPT \cite{GL form factors}, the experimental situation was entirely unclear. We emphasized that there is no way to reconcile some of the published experimental results with the Standard Model. In the meantime, both the accuracy of the theoretical prediction and the quality of the data on the scalar form factor improved considerably, but an experimental discrepancy persists: while the results of ISTRA, KTeV and KLOE are consistent with one another, they disagree with NA48. 

There are not many places where the Standard Model
fails at low energies. Hints at such failures deserve particular attention.  As pointed out by Jan Stern and collaborators \cite{Stern Kmu3}, the Callan-Treiman relation can serve as a probe for the presence of right-handed quark couplings of the $W$-boson -- the existing limits on those couplings are not very stringent. It is premature, however, to interpret the NA48 data as evidence for the occurrence of effects beyond the Standard Model. For the time being, we are merely faced with an experimental discrepancy. The experiment is difficult, because the transition is dominated by the contribution from the vector form factor, while the Callan Treiman relation concerns the scalar form factor. In particular, the radiative corrections must properly be accounted for \cite{Kastner & Neufeld,Neufeld,Bijnens & Ghorbani,Cirigliano Giannotti & Neufeld} . As emphasized by Franzini \cite{Franzini}, the data analysis must cope with very strong correlations between the slopes of the vector and scalar form factors. Also, for the analysis of the data to match their quality, it is essential that the constraints imposed by dispersion theory are respected -- publishing fits based on linear parametrizations of the $t$-dependence or on pole models is meaningless. An analysis of the charged kaon decays collected by NA48 might help clarifying the situation.

\section{Dispersion theory}\label{sec:Dispersion theory}
In the remainder of these lecture notes, I discuss some of the progress made in carrying the low energy analysis to higher energies. First steps in this direction were taken by Gasser and Meissner \cite{Gasser & Meissner}, who compared the representation of the scalar and vector form factors obtained in the framework of \ChPT to two loops with the dispersive representation. In particular, they determined the range of validity of the representation obtained by truncating the chiral perturbation series at one or two loops. 
Indeed, many of the issues discussed in the first part of these lectures involve dispersion theory: the constraints imposed on the form factors by analyticity and unitarity (Watson final state interaction theorem) play a central role in the data analysis. 

In the following, I consider a different example: the $\pi\pi$ scattering amplitude. The chiral perturbation series of this amplitude, even if truncated only at NNLO, is useful only in a very limited range of the kinematic variables -- definitely, the poles generated by $\rho$-exchange are outside this region. The range can be extended considerably by means of dispersion theory, exploiting the fact that analyticity, unitarity and crossing symmetry very strongly constrain the low energy properties of the scattering amplitude.
 
From the point of view of dispersion theory, $\pi\pi$ scattering is
particularly simple: the $s$-, $t$- and $u$-channels represent the same
physical process.  As a consequence, the scattering amplitude
can be represented as the sum of a subtraction term and a dispersion integral over the imaginary part. The subtraction term involves two subtraction constants, which may be identified with the two S-wave scattering lengths. The dispersion
integral exclusively extends over the physical region \cite{Roy}. 

The projection of the amplitude on the partial waves leads to a dispersive
representation for these, the Roy equations. I denote the S-matrix elements by $S^I_\ell=\eta^I_\ell \exp 2 i \delta^I_\ell$ and use the standard normalization for the corresponding partial wave amplitudes $t^I_\ell$:
\be\label{eq:S}S_\ell^I(s)=1+2\,i\,\rho(s)\,t_\ell^I(s)\,,\hspace{2em}
\rho(s)=\sqrt{1-4M_\pi^2/s}\,.\ee The S-matrix elements and the partial wave amplitudes are analytic in the cut $s$-plane. There is a right hand cut ($4M_\pi^2<s<\infty$) as well as a left hand cut ($-\infty<s<0$). The Roy equation for the partial wave amplitude with $I=\ell=0$, for instance, reads 
\bea\label{eq:Roy}
t_0^0(s)=a+(s-4M_\pi^2)b+\sum_{I=0}^2\,\sum_{\ell=0}^\infty\,
\int_{4M_\pi^2}^\infty ds'\, K^{0I}_{0\ell}(s,s')\,\mbox{Im}\hspace{0.1em}
t^{I}_\ell(s') .\eea
As mentioned above, the equation contains two subtraction constants, which can be  expressed in terms of the S-wave scattering lengths:
\be a = a_0^0, \hspace{1cm} b=(2a^0_0-5a^2_0)/12 M_\pi^2.\ee
The kernels $ K^{II'}_{\ell\ell'}(s,s')$ are explicitly known algebraic expressions which only involve the variables $s,s'$ and the mass of the pion, e.g.\\
\be\label{eq:K0000} K_{00}^{00}(s,s')=
\underbrace{ \frac{1}{\pi(s'-s)}}_{r.h.cut} 
+\underbrace{\frac{2\,\ln\{(s+s'\!-4M_\pi^2)/s'\}}
{3\pi(s-4M_\pi^2)}- 
\frac{5s'+2s-16M_\pi^2}{3\pi s'(s'-4M_\pi^2)}}_{l.h.cut}\,.\ee
The integrals on the right hand side of (\ref{eq:Roy}) thus only involve observable quantities: the imaginary parts of the partial waves. 

The pioneering work on the physics of the Roy equations was carried out more
than 30 years ago \cite{MP}. The main problem encountered at that time was that
the two subtraction constants were not known. These dominate the dispersive representation at low energies, but since the data available at the time were consistent with a very broad range of S-wave scattering lengths, the Roy equation ana\-ly\-sis was not conclusive. The insights gained by means of \ChPT thoroughly changed
the situation. Since the S-wave scattering lengths are now known very accurately, the Roy equations have become a very sharp tool for the analysis of the $\pi\pi$ scattering amplitude.

\section{\boldmath Mathematics of the Roy equations}\label{sec:math}

The mathematical properties of the Roy equations are quite remarkable and are discussed in detail in the literature. For an extensive review and references to the original literature, I refer to \cite{ACGL}. In the following, I restrict myself to those features that are essential for an understanding of the consequences of these equations. It is convenient to indicate the isospin of the partial waves as an index: S$^0$, S$^2$ denote the S-waves of isospin 0 and 2, respectively, the P-wave is referred to as P$^1$, while D$^0$, D$^2$ denote the two D-waves etc. 

The Roy equations were derived from axiomatic field theory, for values of $s$ in the interval $-4M_\pi^2<s<60M_\pi^2$  \cite{Roy}. If the scattering amplitude obeys Mandelstam analyticity, then the derivation goes through on a slightly larger domain, $-4M_\pi^2<s<68 M_\pi^2$, but even then, we can make use of these equations only on a finite interval. The upper end of the interval on which the Roy equations are solved is referred to as the matching point. I denote the corresponding value of $s$ by $s_m$. 

If we for the moment treat the imaginary parts of all other partial waves and the two scattering lengths as a given input, the Roy equation (\ref{eq:Roy}) amounts to a representation of the function Re$\hspace{0.05cm}t_0^0(s)$ as an integral over Im$\hspace{0.05cm}t^0_0(s')$ plus a known remainder. In the elastic region, unitarity imposes a second such relation - we thus have two equations for the two unknown functions Re$\hspace{0.05cm}t_0^0(s)$ and Im$\hspace{0.05cm}t^0_0(s)$. Hence no freedom appears to be left: bookkeeping suggests that equation (\ref{eq:Roy}) determines these two functions in terms of the given input.  

If we restrict ourselves to the elastic region ($s_m<16M_\pi^2$), then the naive expectation is indeed correct:  the solution is unique. In order to push the matching point to higher energies, we need to know the elasticity $\eta_0^0(s)$, but if that is the case, we still have two equations for two unknowns.  The properties of the system of equations, however, change if the matching point is pushed up. The solution remains unique only if the phase at the matching point stays below $\frac{1}{2}\pi$. If $\delta_0^0(s_m)$ is in the interval $\frac{1}{2}\pi<\delta_0^0(s_m)<\pi$, the system admits a one-parameter family of solutions, for $\pi<\delta_0^0(s_m)<\frac{3}{2}\pi$, the manifold of solutions is of dimension 2, etc. The available phase shift analyses indicate that the phase $\delta_0^0(s)$ passes through $\frac{1}{2}\pi$ somewhere around 0.85 GeV, goes through $\pi$ in the vicinity of $K\Kbar$ threshold, but stays below $\frac{3}{2}\pi$ on the entire range where the Roy equations are valid. If the matching point is taken at the upper end of this range, equation (\ref{eq:Roy}) thus admits a two-parameter family of solutions \cite{Gasser & Wanders}. 

Analogous statements also hold for the Roy equations obeyed by the other partial waves. For the P-wave, the phase $\delta_1^1(s)$ reaches $\frac{1}{2}\pi$ around the mass of the $\rho$ and remains below $\pi$ for $s<68M_\pi^2$. Accordingly, if the matching point of the P-wave is taken above $M_\rho$, the Roy equation for $t^1_1(s)$ admits a one-parameter family of solutions. The exotic S-wave, $\delta_0^2(s)$, on the other hand, is negative on the entire range where the Roy equations are valid and stays above $-\frac{1}{2}\pi$. In this case, the number of free parameters is equal to -1, irrespective of the choice of the matching point: the Roy equation for $t^2_0(s)$ does not in general admit a solution if the imaginary parts of all other waves (as well as Im$\hspace{0.05cm}t^2_0(s)$ for $s>s_m$) are prescribed arbitrarily. The input must be tuned in order for a solution to exist at all. For all of the higher partial waves, the phase remains below $\frac{1}{2}\pi$ on the interval where the Roy equations hold, so that the solution is unique -- if the phase at the matching point is negative, the input needs to be tuned for a solution to exist. 

The higher waves only play a minor role at low energies. Although, in principle, their properties are correlated with those of the S- and P-waves, we may first use a phenomenological parametrization for the partial waves with $\ell\geq2$ and solve the Roy equations for the S- and P-waves with this input. Then, the Roy equation for the D-, F-, G-waves can be solved, one by one, using the representation for the S- and P-waves  obtained in the first step (on the range where the Roy equations are valid, the contributions from the imaginary parts of the partial waves with $\ell\geq5$ are too small to matter). If the S- and P-waves are known, the Roy equations for the higher waves fix their behaviour at low energies within very small uncertainties. Finally, we may perform an iteration, inserting the representation obtained for the imaginary parts of the waves with $\ell\geq2$ in the Roy equations for the S-and P-waves. There is no need for further iterations because the changes found in these waves are tiny.  

The heart of the matter is a coupled system of three integral equations for the three partial waves $t^0_0(s),t^1_1(s),t^2_0(s)$.  The input of the calculation consists of the following parts: \begin{enumerate}\itemsep = -0.1cm\item  S-wave scattering lengths \item Elasticities of the S-and P-waves below $s_m$\item Imaginary parts of the S- and P-waves above $s_m$ \item Imaginary parts of the higher partial waves\end{enumerate} If the matching point is taken below the $\rho$ mass, the system only admits a solution if this input  is tuned. If the matching point is somewhere in the range between 0.77 and 0.85 GeV (more precisely, the range where $\delta_1^1(s_m)> \frac{1}{2}\pi$, but $\delta_0^0(s_m)< \frac{1}{2}\pi$), the solution is unique. Above, 0.85 GeV, but below $2M_K$, there is a one-parameter family of solutions and for $s_m>4M_K^2$, in particular, if the Roy equations are solved on the entire interval where they are valid, the solution contains two free parameters \cite{Wanders 2000}. In order to arrive at a unique solution, we may, for instance prescribe two of the phases at a suitable energy. It is convenient to use the values of $\delta_0^0(s)$ and $\delta_1^1(s)$ at $\sqrt{s}=0.8\hspace{0.05cm}\mbox{GeV}$ for this purpose. I denote the value of s at this energy by 
\bdm \sA= (0.8\hspace{0.05cm}\mbox{GeV})^2\fs \edm

\section{\boldmath Low energy analysis of $\pi\pi$ scattering}\label{sec:Low energy}
\label{sec:Low energy analysis}
\begin{figure}[thb]
\centering\vspace{-0.3cm}
\includegraphics[width=7.35cm,angle=-90]{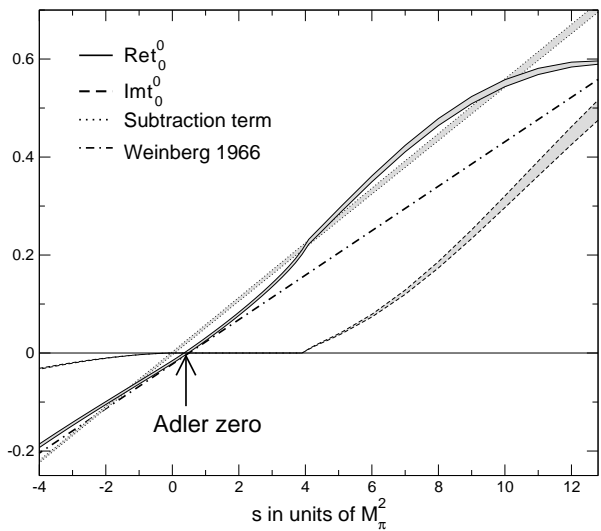}
\caption{\label{fig:subthreshold}Low energy behaviour of the partial wave amplitude $t^0_0(s)$}   
\end{figure}

At low energies, the two S-wave scattering lengths are the most important parameters, because there, the right hand side of the Roy equations is dominated by the subtractions. Figure \ref{fig:subthreshold} demonstrates that the subtraction term in equation (\ref{eq:Roy}) dominates the  behaviour of the partial wave $t^0_0(s)$ throughout the region shown, which extends to about 500 MeV. As predicted by current algebra (tree graphs of the effective theory), $t_0^0(s)$ contains an Adler zero and grows approximately linearly with $s$. 

As discussed in the preceding section, the behaviour in the low energy region is not fully determined by the input listed, but involves two additional low energy degrees of freedom, which can be identified with  $\delta_0^0(\sA)$, $\delta_1^1(\sA)$. The value of $\delta_1^1(\sA)$ is known well, because the Watson final state interaction theorem connects the P-wave phase shift with the vector form factor of the pion, which is accurately determined by the beautiful data on $e^+e^-\rightarrow \pi^+\pi^-$ and $\tau^-\rightarrow\nu_\tau\,\pi^-\pi^0$. The result obtained in \cite{ACGL} reads $\delta_1^1(\sA)=108.9^\circ\pm 2^\circ$. The experimental information about $\delta^0_0(\sA)$, on the other hand, is comparatively meagre -- this currently represents the main source of uncertainty in low energy $\pi\pi$ scattering. In \cite{ACGL}, we observed that phase differences are more easily determined than the phases themselves. Indeed, in contrast to the experimental information about  $\delta^0_0(\sA)$, the one for $\delta^1_1(\sA)-\delta^0_0(\sA)$ is perfectly coherent. Together with the value of $\delta^1_1(\sA)$ quoted above, the experimental information about this difference implies $\delta_0^0(\sA)=82.3^\circ\pm 3.4^\circ$. 

In order to explore the space of solutions, we first fixed the matching point at 0.8 GeV and ignored the predictions of $\chi$PT, treating $a_0^0,a_0^2$ as free parameters \cite{ACGL}. The consequences of the low energy theorems for $a_0^0,a_0^2$ were worked out separately, in \cite{CGL}, where it was shown, for instance, that the scattering lengths and the effective ranges of the S-, P-, D- and F-waves can be calculated within remarkably small uncertainties on this basis. 

One of the sources of uncertainty arises from the "high energy" part of the input: the dispersion integrals extend to infinity -- phenomenology is used to estimate the contributions from the region above the matching point. For the asymptotic properties of the amplitude, we relied on the literature, in particular on the work of Pennington \cite{Pennington}, who had carefully examined the relationship between the behaviour at low energies and the high energy properties of the amplitude. For an update of the Regge parametrization used in this context, see \cite{CCL Beijing}.  It is essential that the Roy equations involve two subtractions, so that  the kernels fall off with the third power of the variable of integration. The left hand cut plays an important role here: as can be seen in equation (\ref{eq:K0000}), the part of the kernel that accounts for the right hand cut falls off only with the first power of the variable of integration, but the high energy tail is cancelled by the contribution from the left hand cut. This ensures that the contributions from the low energy region dominate. 

The results obtained in \cite{ACGL} were confirmed by Stern and collaborators \cite{Stern Roy}. Moreover, these authors made fits to the data available at the time, in order to determine the scattering lengths from experiment and to compare the results with the theoretical predictions. The experimental information about the low energy behaviour of S$^2$ played an important role in their analysis. Unfortunately, this information is not coherent. In particular, the authors had to make a choice between the two phase shift analyses reported in \cite{Hoogland}. With the choice made, the results obtained for $a_0^0$, $a_0^2$ turned out not to be in satisfactory agreement with the theoretical predictions. A similar analysis has now been carried out for the $K_{e4}$ data of NA48/2 \cite{NA48 paper on Ke4,Bloch-Devaux Anacapri 2008} -- as mentioned already in section \ref{sec:Precision experiments}, these data confirm the theory to remarkable precision.

The approach of Yndur\'ain and collaborators \cite{Yndurain,KPY III} is quite different. 
These authors do not make an attempt at solving the Roy equations, but only use these to test and improve their parametrization of the data, making fits where the difference between the left- and right hand sides of the Roy equations and the one between the parametrization and the data is given comparable weight. Indeed, their parametrization underwent a sequence of gradual improvements, which removed some of the deficiencies pointed out in \cite{CCGL,Azores,Lisbon}. In particular, the residue of the $\rho$-trajectory increased step by step, as a consequence of the Olsson sum rule. The representation of the higher partial waves was improved and the experimental determination of the scattering lengths described in \KPYIII \cite{KPY III} is now consistent with our predictions. Unfortunately, however, an important difference to our representation persists: their phase $\delta_0^0(s)$ still contains a kink (discontinuity in the first derivative) at 932 MeV, as well as a hump below that energy. As discussed in detail in \cite{Azores}, these phenomena are artefacts, produced by a parametrization that is not flexible enough. The analysis in \cite{Caprini} corroborates this conclusion.  

Incidentally, the kink and the hump are also responsible for the remaining disagreement in some of the threshold parameters: the contributions from Im\hspace{0.1em}$t^0_0(s)$  to the sum rules for these quantities are not the same (the sum rules are listed in Eqs. (14.1) and (14.3) of \cite{ACGL}). Replacing the parametrization of $\delta_0^0(s)$ below $K\Kbar$ threshold by ours, but taking all other contributions from \KPYIII \cite{KPY III}, the result  reproduces all of the entries listed in table 2 of \cite{CGL}, within errors. In other words, as far as the integrals relevant for the threshold parameters are concerned, the only remaining difference that matters concerns the behaviour of $\delta_0^0(s)$ below $K\Kbar$ threshold. Work aimed at improving the quality of that part of the representation is in progress \cite{Pelaez}.  

\section{\boldmath Behaviour of the S-wave with $I=0$}\label{sec:delta00}
As mentioned in the preceding section, the value of $\delta^0_0(\sA)$ currently represents the main source of uncertainty in low energy $\pi\pi$ scattering. The quoted range, $\delta_0^0(\sA)=82.3^\circ\pm 3.4^\circ$, which follows from the experimental information on the phase difference $\delta_1^1(\sA)-\delta_0^0(\sA)$ and on $\delta_1^1(\sA)$, does not cover all of the data on $\delta_0^0(\sA)$, which are contradictory. While the result of the experiment described in the 1973 PhD thesis of Wolfgang Ochs \cite{Ochs,Hyams}, for instance, is perfectly consistent with the Roy equations and does lead to a value of $\delta_0^0(\sA)$ in this range, the phase shift analysis of the polarized data of the CERN-Cracow-Munich collaboration \cite{Kaminski Lesniak & Rybicki} calls for a higher value. As shown by Kami\'nski, Le\'sniak and Loiseau \cite{Kaminski Lesniak & Loiseau}, the phase ambiguity occurring in that analysis can be resolved by means of the Roy equations. The solutions obtained by these authors are of good quality: the difference between input and output for the real parts are of order $10^{-3}$, for the S-waves as well as for the P-wave. According to figure 2a in \cite{Kaminski Lesniak & Loiseau}, the resulting fit yields  $\delta^0_0(\sA)\simeq 87^\circ$. In view of the relatively large errors attached to the phase shift in \cite{Kaminski Lesniak & Rybicki}, this result must come with a sizable uncertainty and may thus not be inconsistent with the range obtained in \cite{ACGL}, but it is on the high side. 
\begin{figure}[thb]\centering
\includegraphics[height=.5\textheight,angle=-90]{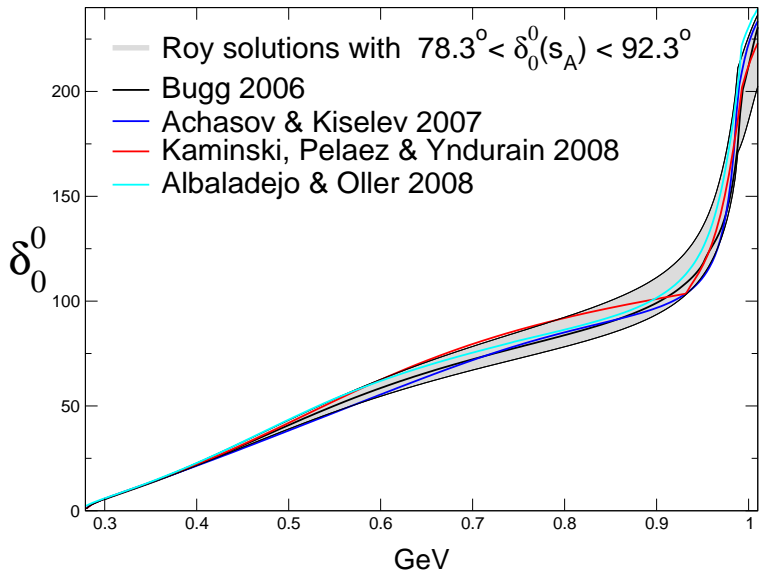}
\caption{\label{fig:delta00}
Behaviour of $\delta_0^0$ below $K\bar{K}$ threshold}  
\end{figure}

The parametrizations of Kami\'nski, Pel\'aez and Yndur\'ain \cite{KPY III} yield even higher values: $\delta^0_0(\sA)=90.7^\circ\pm 0.7$ (A), $\delta^0_0(\sA)=90.5^\circ\pm 0.7$ (B). In view of the remarkably small error, these results disagree with those obtained from $\delta^1_1(\sA)-\delta^0_0(\sA)  $ \cite{ACGL} or from a Roy equation fit to the data of \cite{Ochs}. One of the reasons for arriving at such a high value is that the authors include the result for the phase difference $\delta_0^0(M_K^2)-\delta^2_0(M_K^2)$  obtained from $K\rightarrow \pi\pi$ \cite{Cirigliano Kpipi} in their fitting procedure. This pulls the value of  $\delta_0^0(\sA)$ up. The response of the Roy equations to this change in the input value for $\delta_0^0(\sA)$ is an increase in $\delta_0^0(M_K^2)-\delta^2_0(M_K^2)$ of $2^\circ$. The fit obtained in \KPYIII yields a somewhat larger shift: the value for  $\delta_0^0(M_K^2)-\delta^2_0(M_K^2)$ is $50.9^\circ \pm 1.2^\circ$, higher than our result by $3.2^\circ$. The difference is produced by the kink mentioned in the preceding section, which can also be seen in figure \ref{fig:delta00}. The kink generates a violation of causality and hence of the Roy equations: while our amplitude or the one of Kami\'nski, Le\'sniak and Loiseau \cite{Kaminski Lesniak & Loiseau} do represent decent approximate solutions of the Roy equations, the one in \KPYIII does not: in the region between 0.7 and 1 GeV, the difference between input and output for the real parts of the S-waves is of order 0.1.  Quite irrespective of these details, the increase in the phase difference $\delta_0^0(M_K^2)-\delta^2_0(M_K^2)$ produced by an increase in the value of $\delta_0^0(\sA)$, even a very large one, is much too small to bridge the gap between scattering and decay, which is of the order of $10^\circ$. 
The puzzle discussed in section \ref{sec:scalar meson dominance} needs to be resolved before information about $\pi\pi$ scattering can reliably be extracted from the decay $K\rightarrow \pi\pi$.  

Although the possibility that the phase $\delta_0^0(\sA)$ is above the range obtained from  $\delta^1_1(\sA)-\delta^0_0(\sA)  $ looks unlikely, it cannot be ruled out entirely. For this reason, when analyzing the pole position of the $\sigma$ \cite{CCL}, we stretched the error bar towards higher values and used
\be\label{eq:delta00A} \delta_0^0(\sA)=82.3\, \rule[-0.2em]{0em}{1em}^{+10}_{-\,4}\fs\ee 
The band in figure \ref{fig:delta00} shows the corresponding range of Roy solutions, together with the parametrizations of $\delta_0^0(s)$ proposed in \cite{KPY III,Bugg,Achasov,Albaladejo}. The plot shows that the range (\ref{eq:delta00A}) covers the values of $\delta_0^0(\sA)$ obtained with these.

\section{Pole formula}\label{sec:Pole formula}
The positions of the S-matrix poles represent universal properties of QCD, which are unambiguous even if the width of the resonance turns out to be large, but they concern the non-perturbative domain, where an analysis in terms of the local degrees of freedom -- quarks and gluons -- is not in sight. One of the reasons why the values for the pole position of the $f_0(600)\equiv \sigma$ quoted by the Particle Data Group cover a very broad range is that all but one of these either rely on models or on the extrapolation of simple parametrizations: the data are represented in terms of suitable functions on the real axis and the position of the pole is determined by continuing this representation into the complex plane. If the width of the resonance is small, the ambiguities
inherent in the choice of the parametrization do not significantly affect the
result, but the width of the $\sigma$ is not small. For a thorough discussion of the sensitivity of the pole position to the freedom inherent in the choice of the parametrization, I refer to \cite{Caprini}.

The determination of the $\sigma$ pole provides a good illustration for the strength of the dispersive method and for the relative importance of the various terms on the right hand side of the Roy equations. Using known results of general quantum field theory \cite{Martin book,Roy Wanders 1978}, we have shown that the Roy equations also hold for complex values of $s$, in the intersection of the relevant Lehmann-Martin ellipses \cite{CCL}. 
\begin{figure}[thb]\vspace{-0.5cm}\centering
\includegraphics[width=6cm,angle=-90]{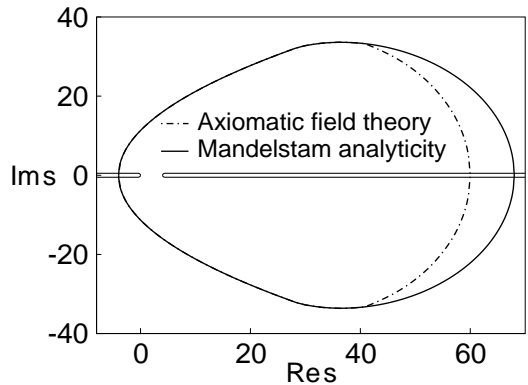}
\caption{\label{fig:Potato}Domain of validity of the Roy equations ($s$ in units of $M_\pi^2$)}
\end{figure}
The pole sits on the second sheet, which is reached from the first by analytic continuation from the upper half plane into the lower half plane, crossing the real axis in the interval $4M_\pi^2<s< 16 M_\pi^2$, where the scattering is elastic. For a real value of $s$ on this interval, we have $S_\ell^I(s\pm i\,\epsilon)^I=\exp \pm 2\, i\,\delta^I_\ell(s)$, so that $S_\ell^I(s+i\,\epsilon)^{II}=S_\ell^I(s-i\,\epsilon)^I=1/S^I_\ell(s+i\,\epsilon)^I $. Hence the relation 
\be S^I_\ell(s)^{II}=1/ S^I_\ell(s)^{I}\ee 
holds on a finite interval of the real axis. Since the equation connects two meromorphic functions, it also holds for complex values of $s$. A pole on the second sheet thus occurs if and only if $S^I_\ell(s)$ has a zero on the first sheet. The net result of this discussion is that we have an exact formula for resonances:
\be S^I_\ell(s)=0\fs\ee 
For resonances with the quantum numbers of the vacuum, $I=\ell=0$, the element $S^0_0(s)$ of the S-matrix is relevant. It is specified explicitly in equations (\ref{eq:S}) and (\ref{eq:Roy}), in terms of the imaginary parts of the partial waves on the real axis. The formula thus exclusively involves observable quantities and can be evaluated for complex values of $s$ just as well as for real values. It provides for the analytic continuation necessary to determine the resonance position -- a parametrization is not needed for this purpose. 

\section{The lowest resonance of QCD}\label{sec:Lowest resonance}
Inserting our central representation for the scattering amplitude in (\ref{eq:Roy}), we find that, in the region where the Roy equations are valid, the function $S^0_0(s)$ has two zeros in the lower half of the first sheet: one at $\sqrt{s}= 441 - i\, 272$ MeV, the other in the vicinity of 1 GeV \cite{CCL}. While the first corresponds to the $\sigma$, the second zero represents the well-established resonance $f_0(980)$. Our analysis sheds little light on the properties of the latter, because the location of the zero is sensitive to the input used for the elasticity $\eta^0_0(s)$ -- the shape of the dip in $\eta^0_0(s)$ and the position of the zero represent two sides of the same coin.  For this reason, I only discuss the $\sigma$.

We are by no means the first to find a resonance in the vicinity of the above position. In the list of papers quoted by the Particle Data Group \cite{PDG 2007}, the earliest one with a pole in this ball park appeared more than 20 years ago \cite{Beveren}.  What is new is that we can perform a controlled error calculation, because our method is free of the systematic theoretical errors inherent in models and parametrizations. For this purpose, it is convenient to split the right hand side of the Roy equation for $t^0_0(s)$ into three parts:
\begin{description}
\itemsep = -0.05cm
\item a. Subtraction constants
\item b. Contribution from Im\hspace{0.1em}$t^0_0(s)$ below $K\Kbar$ threshold
\item c. Contributions from higher energies and other partial waves
\end{description}

\vspace{0.1cm}
\noindent{\bf a.\ Subtraction constants}

\vspace{0.2cm} The subtraction constants are determined by the S-wave scattering lengths. The predictions for these read $a_0^0=0.220\pm 0.005$ and $a^2 _0=-0.0444\pm 0.0010$ \cite{CGL}. Following error propagation, we find that an increase in $a_0^0$ by 0.005 shifts the pole position by $(-2.4 +i\,3.8)$ MeV, while the response to an increase in $a_0^2$ by 0.0010 is a shift of $(0.8 -i\, 4.0)$ MeV \cite{CCL}. These numbers show that the error in the pole position due to the uncertainties in the subtraction constants are small.

\vspace{0.2cm}
\noindent {\bf b.\ Contribution from Im\hspace{0.1em}$t^0_0(s)$ below $K\Kbar$ threshold}

\begin{figure}[thb]\centering
\includegraphics[width=6cm,angle = -90]{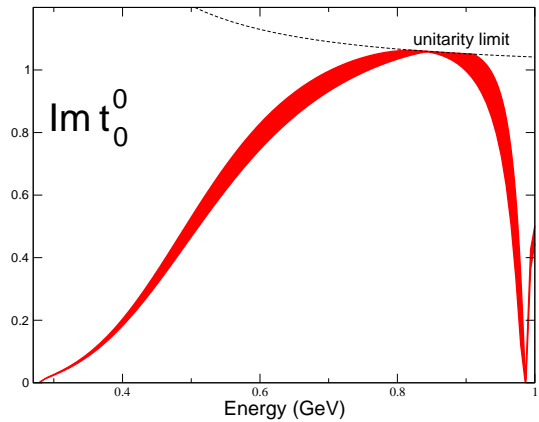}
\caption{\label{fig:imt00}
Behaviour of Im$\hspace{0.05cm}t_0^0$ below $K\bar{K}$ threshold}  
\end{figure}

\vspace{0.2cm}Below $K\Kbar$ threshold, the S-waves are elastic to a very good approximation. As shown in figure \ref{fig:imt00}, the function Im\hspace{0.1em}$t^0_0(s)$ shows a broad bump, nearly hits the unitarity limit somewhere between 800 and 900 MeV and then rapidly drops, because the phase steeply rises, reaching 180$^\circ$ in the vicinity of 2 $M_K$. Hence there is a pronounced dip in Im\hspace{0.1em}$t^0_0(s)$ near $K\Kbar$ threshold. The bump seen in the imaginary part in figure \ref{fig:imt00} may be viewed as a picture of the broad resonance we are discussing here. 

Below $K\Kbar$ threshold, the behaviour of the imaginary part is controlled almost entirely by the phase shift $\delta^0_0(s)$. Replacing the integral over our central representation for Im\hspace{0.1em}$t^0_0(s)$ from $4M_\pi^2$ to $4 M_K^2$ by the one of Bugg  \cite{Bugg}, but leaving everything else as it is, the pole moves to 444 - i 267 MeV. Repeating the exercise with the representations of Kami\'nski, Pel\'aez and Yndur\'ain \cite{KPY III}, Achasov and Kiselev \cite{Achasov} and Albaladejo and Oller  \cite{Albaladejo}, the pole is shifted to 458  - i 253 MeV, 438 - i 274 MeV and  451 - i 257 MeV, respectively. If the imaginary part of $t^0_0(s)$ is evaluated with the lower edge of the band shown in figure \ref{fig:delta00}, which corresponds to $\delta_0^0(\sA)=78.3^\circ$, the pole occurs at 435 - i 276 MeV, while for the upper edge of the band, characterized by $\delta_0^0(\sA)=92.3^\circ$, the pole sits at 456 - i 262 MeV.  

\vspace{0.3cm}
\noindent {\bf c.\ Contributions from higher energies and other partial waves}

\vspace{0.2cm} Finally, I turn to the contributions of the third category: higher energies and other partial waves. Among these, the one from the P-wave, for example, is by no means negligible, but, as mentioned above, this wave is known very well. In fact, in the vicinity of the zero of $S^0_0(s)$, the sum of the contributions of this category can be worked out quite accurately. The net uncertainty in the pole position from this source is $\pm 4 \pm \mbox{i}\, 6$ MeV. As a check, we can simply replace our central representation for the contributions of category c by the one in \cite{KPY III}, retaining our own representation only for the remainder. The operation shifts the pole position by $ - 0.6 -\mbox{i}\, 1.2$ MeV, well within the estimated range.

Adding the errors up in square, the result for the pole position becomes
\be\label{eqmsigma} \sqrt{s_\sigma}=441\, \rule[-0.2em]{0em}{1em}^{+16}_{-\,8}-
\,\mbox{i}\;272\,\rule[-0.2em]{0em}{1em}^{+\,9}_{-12.5}\;\mbox{MeV}\;\; \cite{CCL}.\ee 
The error bars account for all sources of uncertainty and are an order of magnitude smaller than for the estimate $\sqrt{s_\sigma}$ = (400 - 1200) - i (250 - 500) MeV quoted by the Particle Data Group \cite{PDG 2007}: the position of the lightest resonance of QCD can now be calculated reliably and quite accurately.
\section*{Acknowledgement}
I thank Irinel Caprini, Gilberto Colangelo, J\"urg Gasser and Emilie Passemar for a very pleasant collaboration and Kolia Achasov, Miguel Albaladejo, Vincenzo Cirigliano, Hans Bijnens, David Bugg, Gerhard Ecker, Paolo Franzini, Shoji Hashimoto, Daisuke Kadoh, Robert Kami\'nski, Takashi Kaneko, Lesha Kiselev, Yoshinobu Kuramashi, Helmut Neufeld, Jos\'e Antonio Oller, Jos\'e Pel\'aez, Toni Pich and Chris Sachrajda  for informative discussions and correspondence. Also, it is a pleasure for me to thank Nino Zichichi for a most enjoyable stay at Erice.

\end{document}